\documentclass{emulateapj}
\usepackage{graphicx, amssymb, amsthm}

\makeatother

\newcommand\etal{{\it et~al.~}}

\begin{document}
\title{Ionization Front Instabilities in Primordial H II Regions}
\author{Daniel Whalen\altaffilmark{1} \& Michael L. Norman\altaffilmark{2}}
\altaffiltext{1}{Applied Physics (X-2), Los Alamos National
Laboratory}
\altaffiltext{2}{Center for Astrophysics and Space Sciences,
University of California at San Diego, La Jolla, CA 92093, U.S.A.
Email: dwhalen@cosmos.ucsd.edu}

\begin{abstract} 

Radiative cooling by metals in shocked gas mediates the formation of ionization front instabilities 
in the galaxy today that are responsible for a variety of phenomena in the interstellar medium, from the 
morphologies of nebulae to triggered star formation in molecular clouds.  An important question in 
early reionization and chemical enrichment of the intergalactic medium is whether such instabilities 
arose in the H II regions of the first stars and primeval galaxies, which were devoid of metals.  We 
present three-dimensional numerical simulations that reveal both shadow and thin-shell instabilities 
readily formed in primordial gas.  We find that the hard UV spectra of Population III stars broadened 
primordial ionization fronts, causing H$_2$ formation capable of inciting violent thin-shell instabilities 
in D-type fronts, even in the presence of intense Lyman-Werner flux.  The high post-front gas temperatures
associated with He ionization sustained and exacerbated shadow instabilities, unaided by 
molecular hydrogen cooling.  Our models indicate that metals eclipsed H$_2$ cooling in I-front 
instabilities at modest concentrations, from 1 $\times$ 10$^{-3}$ - 1 $\times$ 10$^{-2}$ Z$_\odot$.  
We conclude that ionization front instabilities were prominent in the H II regions of the first stars 
and galaxies, influencing the escape of ionizing radiation and metals into the early universe.

\end{abstract}


\section{Introduction}

Instabilities of several varieties are known to form in ionization fronts (I-fronts) in the galaxy
today, giving rise to structures like elephant trunks and bright rims in star-forming regions of the 
interstellar medium (ISM).  Two prominent types are shadow instabilities \citep{rjw99} and thin-shell 
instabilities \citep{gsf96}, which usually manifest as fingers of ionized gas protruding ahead of the front.
Shadow instabilities form when a density fluctuation is advected through an R-type front, creating 
a dimple that first elongates but then stabilizes.  When the front becomes D-type, the dimple can 
erupt in a jet if the shocked gas radiatively cools and forms a cold dense shell liable to 
fragmentation.  Radiation escapes through fissures in the shell, causing the dimple to fracture and 
jet forward.  Thin-shell instabilities arise when a D-type front encounters a density perturbation.  
If the prefront shocked gas efficiently cools, the perturbation triggers a Vishniac instability 
\citep{v83} in the shell; if not, the shell may instead fracture due to Rayleigh-Taylor instabilities \citep{mn93}.  
Either way, radiation rapidly destabilizes the perturbations by escaping through cracks in the shell and 
driving ionized extensions of gas much larger than the original disturbance into the surrounding medium.  
In a previous paper, we examined the topology of both types of instabilities in the ISM \citep{wn07}, 
confirming that their violence was determined by the cooling efficiency of the shocked gas.

H II regions appeared at the redshifts of first structure formation \citep{wan04,ket04}, beginning 
with the formation of isolated massive stars in the first dark matter halos of mass $\gtrsim$ 1.0 $\times$ 
10$^{5}$ M$\odot$ at $z \sim$ 20 - 30 \citep{abn00,abn02,bcl99,bcl02} and continuing with the rise of the 
first stellar populations in primeval galaxies a few hundred Myr later.  Whether I-front instabilities 
could form in  
pristine H and He at high redshifts is a key question in early cosmological reionization and 
structure formation for several reasons.  First, they may have regulated UV escape from the first 
stars and galaxies, which has risen sharply in importance with the recent discovery of substantial free
electron optical depths at by the \textit{Wilkinson Microwave Anisotropy Probe} (WMAP) 
\citep{spet03,ket03,pet06}, implying that cosmological reionization began much earlier than previously expected.
Second, I-front instabilities may have created density condensations prone to clumping and later 
collapse, forming new generations of stars.  Third, the mixing of metals from the first supernovae  
with remnants of I-front instabilities in relic H II regions determined the extent to which the 
early intergalactic medium (IGM) was chemically enriched.  Contamination of instability relics
may fracture and collapse them on mass scales very different from the stars that created them
\citet{mbh03,yet07}, abruptly splitting the primordial initial mass function (IMF) into a low-mass 
branch in just one generation.  On the other hand, if saturated I-front instabilities drive 
turbulent flows, they may have supported clouds in minihalos or protogalaxies from collapse, 
postponing rather than promoting star formation \citep{wn07}.  Finally, recent work by 
\citet{miz05,miz06} suggests that ionization fronts penetrating molecular cloud cores in the ISM 
may develop unstable modes that reach the centers of the core before the front.  This may disrupt 
collapse of cloud cores into new stars, both in the galaxy and at high redshifts.

Given the central role of radiative cooling in modulating I-front instabilities in the ISM, one 
might question whether they arose at all in primordial ionized flows, in which the atomic or 
molecular hydrogen cooling is far less efficient. Cosmological ionization fronts also differed 
from their galactic counterparts in being driven by the hard UV and soft X-ray spectra of Pop III 
stars and 
miniquasars.  Such spectra exhibit a range of photon mean free paths that can, depending on the 
ambient density, greatly broaden the front, in some cases to as much as the Hubble distance.  It 
is unclear what impact the broadening of the front would have on the formation and evolution of 
unstable modes in I-fronts; the temperature precursor preceding such fronts may preempt their 
formation entirely.

We have performed three-dimensional radiation hydrodynamical calculations of ionization front 
instabilities in which multifrequency radiative transfer is coupled to nine species primordial
chemistry.  We explore the formation of thin-shell and shadow instabilities in both primordial
and incrementally-enriched gas in idealized geometries to disentangle their evolution from 
other features imprinted on the front by density imhomogeneities in real cosmological objects.  
As primordial fronts exit minihalos and protogalaxies they will encounter structures on several
spatial scales that shape the global morphology of the final H II region (e.g., \citet{awb06}).  
For example, halos hosting Population III (Pop III) star formation exhibit filamentary inflows 
and other departures from sphericity on 100 pc scales \citep{abn00,fc00,yet03}.  However, 
protostellar infall likely settles into an accretion disk on 1000 AU scales, collimating the 
emergent front into biconical flows above and below the plane of the disk \citep{tm04}.  Transverse gas 
velocities inherited from the angular momentum of the parent dark matter halo further complicate 
breakout of the nascent front on intermediate scales.  We survey the formation of instabilities 
in primordial fronts in advance of multiscale simulations of UV breakout from primeval stars and
and galaxies. 

In $\S$ 2 we describe the multifrequency upgrades to our photon-conserving radiative transfer in the 
ZEUS-MP reactive flow physics code documented in \citet{wn06}.  Primordial cooling processes are examined in $\S$ 
3, particularly the role of H$_2$ cooling in cosmological ionization fronts.  In $\S$ 4 we present
our numerical models of D-type I-front instabilities due to thin-shell modes with atomic and 
molecular cooling.  We follow the evolution of shadow instabilities associated with Pop III 
blackbody spectra in $\S$ 5, pinpoint the threshold at which metal line cooling triggers thin-shell 
instabilities in D-type fronts in $\S$ 6, and conclude in $\S$ 7. 

\section{Radiative Transfer: Conservative Multifrequency Photon Transport} \label{sect:mnu}

The monoenergetic radiative transfer in our previous studies does not capture postfront gas 
temperatures from first principles or the natural broadening of fronts by hard UV photons.  It 
also neglects molecular hydrogen dissociation by Lyman-Werner photons and photodetachment of H$^-$, 
key processes that regulate the abundance of H$_2$, an important coolant in primordial gas.  Here 
we describe modifications to our numerical method \citep{wn06} for computing radiative rate 
coefficients with a multifrequency algorithm that preserves photon conservation.   

Constraining the number of absorptions in a zone due to all processes to equal the number of photons 
entering the cell minus those exiting leads to
\begin{equation}
n_{abs} \propto 1 \: - \: e^{- \tau}, \vspace{0.075in} \label{nabs}
\end{equation}
where \vspace{0.075in}
\begin{equation}
\tau = \sum_{i=1}^{n} \sigma_i n_i \Delta r, \vspace{0.075in}
\end{equation}
is the total optical depth of the cell (the sum is over all removal processes).  Equations 1 and 2 
are true for a single energy and therefore hold if summed over all energies.  The total absorption 
rates n$_{abs}$ are necessary to determine the flux that reaches the next zone.  On the other hand,  
the rate for a single removal process is a function of an attenuation factor that does not involve 
any of the other types of absorptions:\vspace{0.075in}
\begin{equation}
n_{i} \propto 1 \: - \: e^{- \tau_i}, \vspace{0.075in} \label{ki}
\end{equation}
where $\tau_{i}$ $=$ n$_{i} \sigma_{i} \Delta$r, the optical depth to that interaction.  

The sum of the individual rates evaluated in this manner is inconsistent with n$_{abs}$, but the 
two can be reconciled by enforcing \vspace{0.075in}
\begin{equation}
n_{i} \propto \displaystyle \frac{1 \: - \: e^{- \tau_i}}{\sum_{i=1}^{n}1 \: - \: 
e^{- \tau_i}} \, n_{abs}. \vspace{0.075in}
\end{equation}   
This formulation guarantees that the sum of the rates matches the total rate associated with the 
optical depth of the zone (thus preserving photon conservation) while properly binning individual 
reactions according to their numbers.  We evaluate rate coefficients (where k$_i$ $=$ 
n$_i$/(nV$_{cell}$), where $n$ is the gas number density) at every frequency, assuming the photon 
emission rate $\dot{n}$($\nu_i$) at that frequency, and sum them to obtain the total coefficient 
for the entire spectrum.  We increment energy release in the gas one frequency at a time for a given 
photoreaction as well.  In tests spanning 40 to 2000 logarithmically-spaced bins with a 1 $\times$ 
10$^5$ K Pop III blackbody spectrum, we obtain good convergence with 80 bins above the ionization limit 
of H.

We partition the emission rate ${\dot{n}}_{\gamma}$ of ionizing photons of a Pop III star above from 
\citet{s02} into frequency intervals according to the number rate of a blackbody: \vspace{0.06in}
\begin{equation}
{\dot{n}}_{\gamma} \,=\, \sum_{i=1}^{n}\dot{n}(\nu_i)\Delta\nu \,\propto\, \sum_{i=1}^{n}\displaystyle 
\frac{(\nu_i/c)^2}{e^{h\nu_i/kT}-1}\,\Delta\nu,\vspace{0.06in} 
\end{equation}
where the sum is over h$\nu$ $>$ 13.6 eV.  The emission rate for a particular bin is then \vspace{0.06in}
\begin{equation}
{\dot{n}}_{\gamma,j}\, = \,  \dot{n}(\nu_j)\, \Delta\nu\, = \, \displaystyle \frac{\displaystyle 
\frac{(\nu_j/c)^2}{e^{h\nu_j/kT}-1}}{\sum_{i=1}^{n}\displaystyle \frac{(\nu_i/c)^2}{e^{h\nu_i/kT}-1}}
\,{\dot{n}}_{\gamma}. \vspace{0.06in}
\end{equation}
The procedure is similar for a quasar power-law flux spectrum F($\nu$)d$\nu$ $\propto$ $\nu^{-\alpha}
$d$\nu$.  If $\dot{n}_{QSO}$ is the ionizing emission rate for the quasar, a simple integration over 
frequency yields\vspace{0.06in} 
\begin{equation}
\dot{n}(\nu)\, = \dot{n}_{QSO} \, \displaystyle \frac{{\nu_{th}}^{\alpha}\alpha}{\nu^{\alpha+1}} 
\vspace{0.06in}
\end{equation}
from which follows
\begin{equation}
\dot{n}_{\gamma,j}\, = \, \dot{n}_{QSO} \left\{\sum_{i=1}^{n}\displaystyle \frac{{{\nu}_j}^{\alpha+1}}
{{{\nu}_i}^{\alpha+1}}\right\}^{-1} \vspace{0.06in}
\end{equation}
We truncate the ionizing UV spectrum at 90 eV for Pop III stars.

\begin{deluxetable}{lll}
\tabletypesize{\scriptsize}
\tablecaption{ZEUS-MP Radiative Reaction Processes \label{tbl-1}}
\tablehead{
\colhead{Rate} & \colhead{Reaction} & \colhead{Energy}}
\startdata
  k$_{24}$   &    H + $\gamma$ $\rightarrow$ H$^+$ + e$^-$          &  h$\nu$ $>$ 13.6 eV              \\
  k$_{26}$   &    He$^+$ + $\gamma$ $\rightarrow$ H$e^{++}$ + e$^-$ &  h$\nu$ $>$ 54.4 eV              \\
  k$_{25}$   &    He + $\gamma$ $\rightarrow$ H$e^+$ + e$^-$        &  h$\nu$ $>$ 24.6 eV              \\
  k$_{27}$   &    H$^-$ + $\gamma$ $\rightarrow$ H + e$^-$          &  h$\nu$ $>$ 0.755 eV             \\
  k$_{28}$   &    H$_2^+$ + $\gamma$ $\rightarrow$ H$^+$ + H        &   2.65 eV $<$ h$\nu$ $<$ 21 eV   \\
  k$_{29}$   &    H$_2$ + $\gamma$ $\rightarrow$ H$_2^+$ + e$^-$    &  h$\nu$ $>$  15.42 eV            \\
  k$_{30}$   &    H$_2^+$ + $\gamma$ $\rightarrow$ 2H$^+$ + e$^-$   &     30 eV $<$ h$\nu$ $<$ 70 eV   \\
  k$_{31}$   &    H$_2$ + $\gamma$ $\rightarrow$ 2H                 &  11.18 eV $<$ h$\nu$ $<$ 13.6 eV \\
\enddata
\end{deluxetable}

The radiative processes now in the code are listed in Table \ref{tbl-1}.  We adopted cross sections
from \citet{o89} to calculate k$_{24}$ and k$_{26}$, from \citet{aet97} to compute k$_{27}$ and from 
\citet{sk87} to evaluate k$_{25}$, k$_{28}$, k$_{29}$, and k$_{30}$.  The rate coefficients 
k$_{27}$ and k$_{28}$ both extend below 13.6 eV; we allocate an additional 40 bins equally spaced 
between 0.755 eV and 13.6 eV to evaluate them.  We obtain binned source rates over this energy range
by constructing blackbody number spectra, normalizing them by total stellar luminosities taken from 
\citet{s02} instead of by total ionizing photon emission rates as before.  
Considerable speedups of the radiative transfer ($\sim$ 30\%) can be realized if the linear approximations 
to equations \ref{nabs} and \ref{ki} are adopted in the limit of small optical depths, which are common
for species like H$^-$ and H$_2^+$.  The k$_{31}$ calculation is decoupled from the radiative transfer 
and computed using the H$_2$ self-shielding functions of \citet{db96} corrected for thermal Doppler 
broadening (see also \citet{aet97})\vspace{0.06in}
\begin{equation}
k_{31} \,=\, 1.1 \times 10^8 \bar{F}(\nu) F_{shield},\vspace{0.06in}
\end{equation}
where F$_{shield}$ is obtained from eq.37 of \citet{db96} and $\bar{F}(\nu)$ is the specific flux in s$^{-1}$
ergs cm$^{-2}$ Hz$^{-1}$\vspace{0.06in}
\begin{equation}
\bar{F}(\nu) \,=\, \displaystyle \frac{1}{\Delta\nu} \int F(\nu) \,d\nu\vspace{0.06in}
\end{equation}
and the frequency range corresponds to the Lyman-Werner band 11.18 eV to 13.6 eV.  Total stellar luminosities
are applied to derive F($\nu$) for the blackbody spectrum.  To approximate the effect of gas motion we assign 
a temperature of 10,000 K to the Doppler correction to the shielding function, as in \citet{as07}.  The 
viability of thermal broadening as a proxy for flows in I-fronts is uncertain, but including it reduces 
shielding at intermediate column densities, setting lower limits on H$_2$ cooling in the gas.  

\begin{figure}
\resizebox{3.45in}{!}{\includegraphics{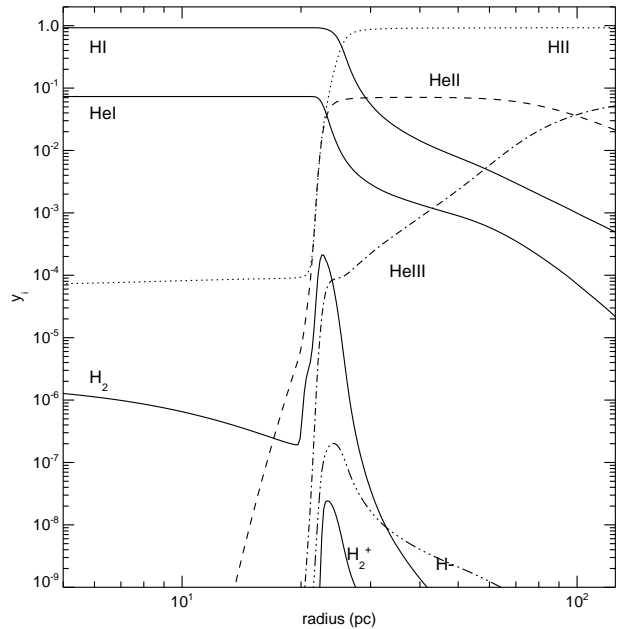}}
\caption{Primordial H and He abundances in a cosmological ionization front at t $=$ 0.6t$_\ast$.  In this 
figure, the radiation wave approaches from the right toward the halo centered at the left boundary.  Compare 
to Fig. 8 of \citet{as07} and the mirror image of Fig. 3 of \citet{rs01}.  Each curve is labeled by its 
species.} \vspace{0.25in}
\label{fig:TIS}
\end{figure}

To test the coupling of the multifrequency transport to the nine-species reaction network \citep{anet97}, we 
simulated the photoevaporation of a cosmological minihalo 540 pc from a 120 M$\odot$ Pop III star .  A 
spherically-symmetric truncated isothermal sphere (TIS) density profile was centered at the left $x$ face
of a 125 pc cartesian box with a uniform mesh of 512 zones.  A geometrically attenuated plane wave propagated 
along the negative x-axis, engulfing the halo.  This is the test case shown in Fig. 8 of \citet{as07}, and 
we adopted identical initial conditions for comparison to their one-dimensional lagrangian radiation hydrocode. 
Number fraction abundances for the primordial H and He species appear in Fig. \ref{fig:TIS} at 0.6t$_\ast$,
where t$_\ast$ is the lifetime of the star.

\begin{figure*}
\epsscale{1.}
\plottwo{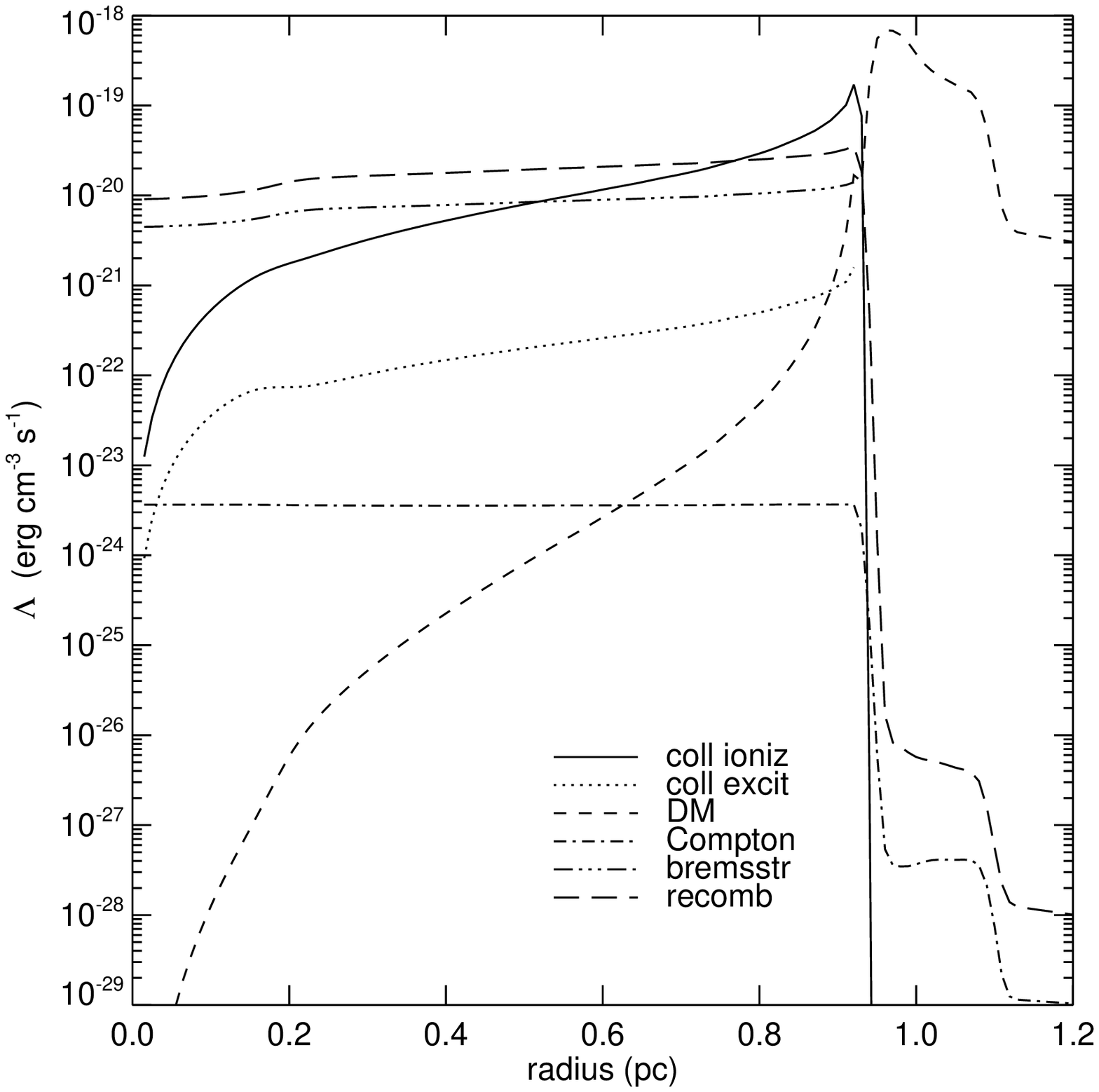}{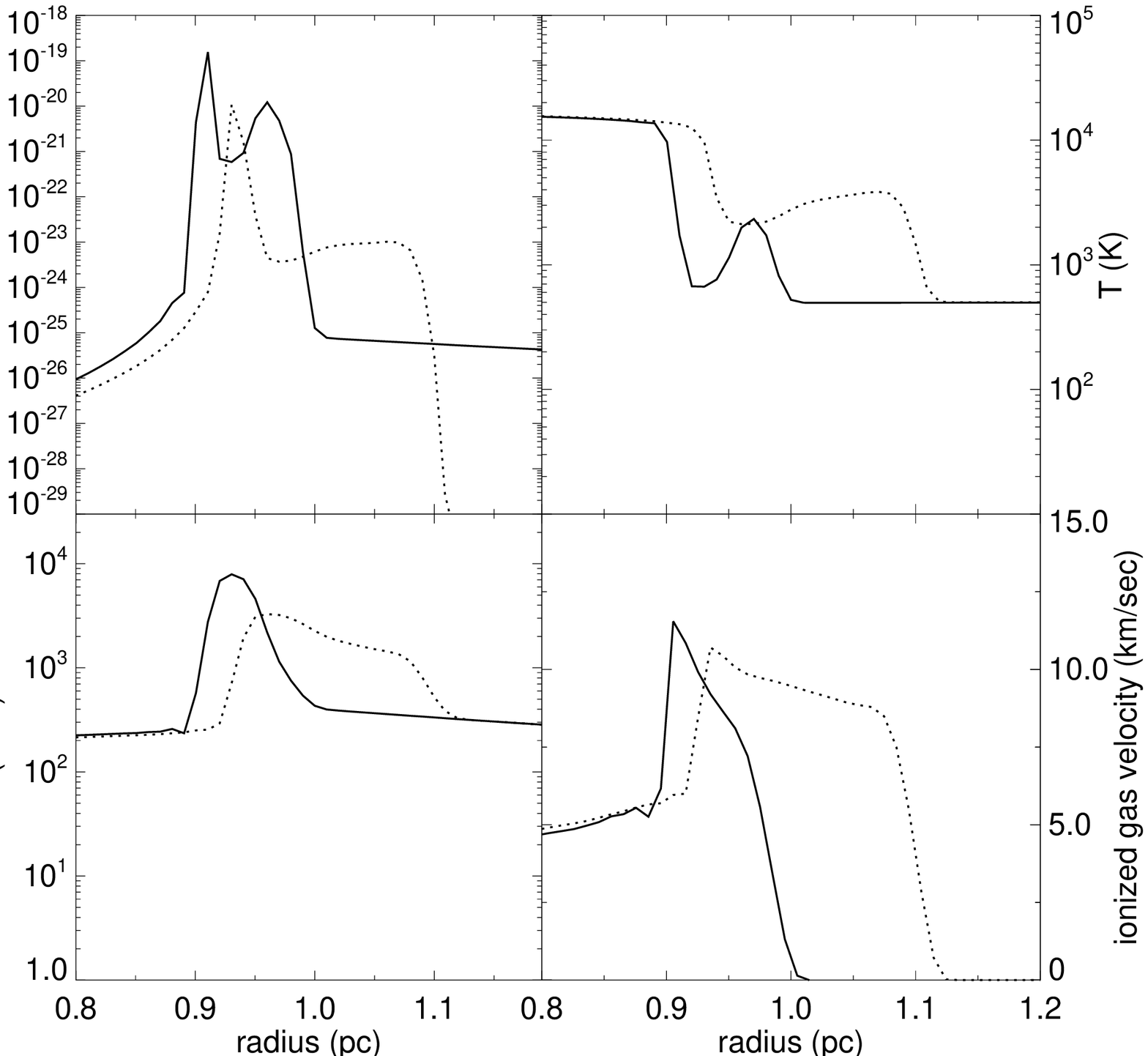}
\caption{Left: Primordial cooling rates in the ionized gas and neutral shocked shell of an I-front at 100 Kyr
not cooled by H$_2$.  Dalgarno \& McCray cooling rates, included for comparison, dwarf atomic H radiative rates 
in the dense shell.  Right:  H$_2$ cooling rate, density, temperature, and velocity profiles for a primordial 
ionization front with no Lyman-Werner photodissociation (solid) and Lyman-Werner photodissociation (dotted).
\label{fig:cool}}  
\vspace{0.1in}
\end{figure*}

Perhaps the most striking feature is the shell of molecular hydrogen created in the partially ionized outer
layers of the front \citep{rs01} (notice its peak coincides with an H II fraction $\sim$ 0.1).  The hard UV 
photons in the spectrum transit a range of mean free paths in the neutral gas, broadening the front to 
approximately 5 pc.  Free electron abundances at temperatures of a few thousand K in the finite width of the 
front rapidly catalyze H$_2$ production by the H$^-$ and H$_2^+$ channels: \vspace{0.1in}
\begin{equation}
H + e^- \rightarrow H^- + \gamma \hspace{0.25in} H^- + H \rightarrow H_2 + e^- 
\end{equation}
\begin{equation}
H + H^+ \rightarrow H_2^+  + \gamma \hspace{0.25in} H_2^+ + H \rightarrow H_2 + H^+ \vspace{0.1in}
\end{equation}
The peak H$^-$ and H$_2^+$ abundances are in step with the H$_2$ peak, as expected since they are the key 
intermediaries for molecular hydrogen production in these relatively low densities.  The H$^-$ channel is 
dominant because it is more abundant and it has much higher H$_2$ formation rates in these regimes.  Fig. 
\ref{fig:TIS} is in good agreement with Fig. 8 of \citet{as07} and is similar to Fig. 3 of \citet{rs01}, 
which depicts H and He abundances for a primordial ionization front in a static uniform medium at the mean 
cosmological density (their figure should be compared to the mirror image of ours because their front propagates 
to the right).  We note the relative abundances of H$^-$ and H$_2^+$ to be a general feature of partially
ionized primordial gas at a few thousand K at these densities (compare Fig. 3 in \citet{anet97}, which features 
collisionally ionized gas collapsing in a pancake shock).

\section{Shock Cooling in Primordial Gas}

\begin{figure*}
\epsscale{1.1}
\plotone{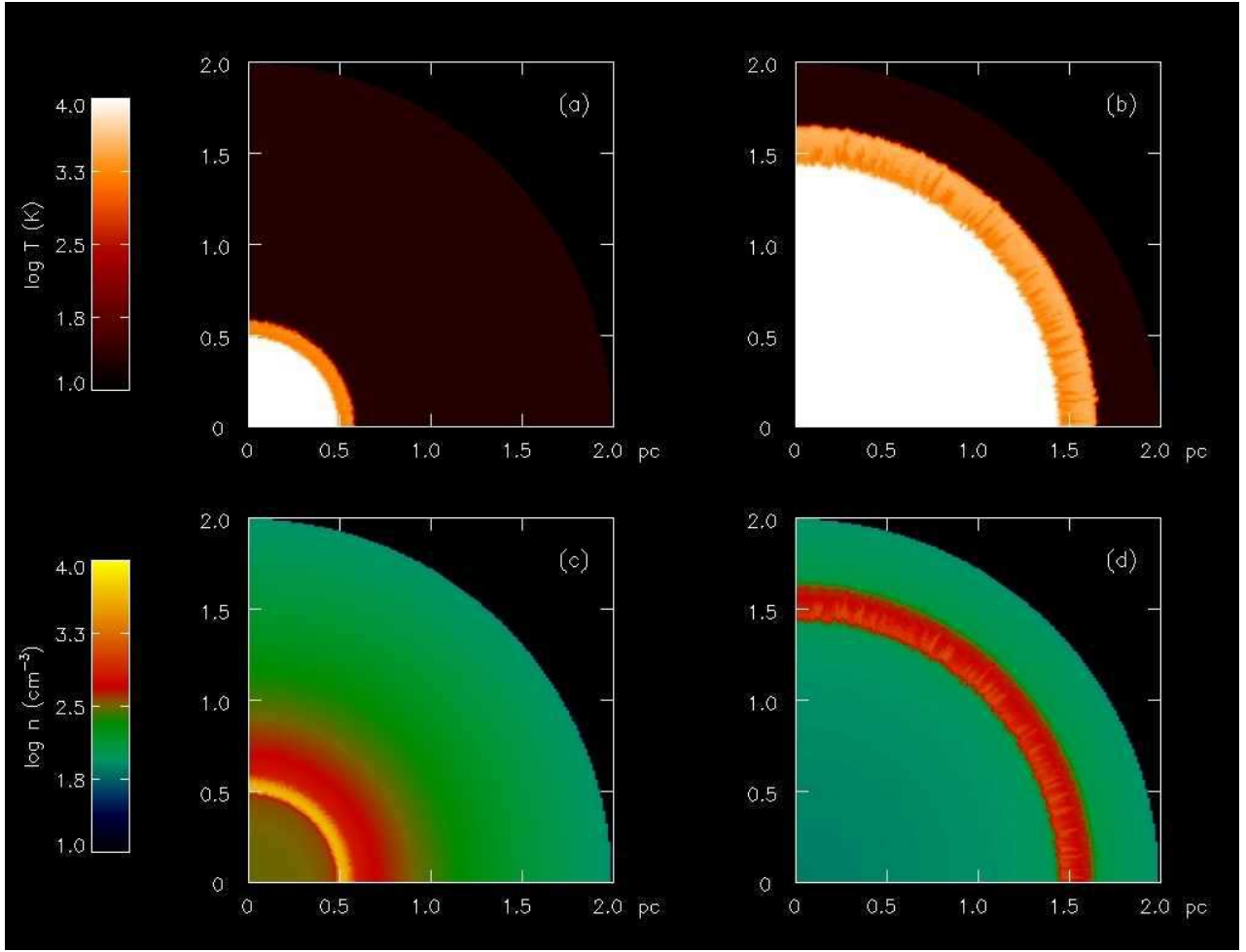}\vspace{0.15in}
\caption{Evolution of a D-type front in an primordial r$^{-2}$ envelope.  Panels (a) and (b):  temperatures
at 47.6 kyr and 135 kyr, respectively.  Panels (c) and (d):  densities at 47.6 kyr and 135 kyr, respectively.
\label{fig:S31PM}} 
\vspace{0.075in}
\end{figure*}
Five mechanisms that cool primordial H and He in I-front shocks are electron collisional and ionizational 
cooling, recombinational cooling, bremsstrahlung, and Compton cooling (primarily in the early universe).  
The free electrons required for these processes are present in minute quantities (n$_e$/n$_H$ $\sim$ 1 
$\times$ 10$^{-4}$) at high redshifts and are either remnants of the recombination era or created in 
cosmological shocks.  These channels cool H and He far less efficiently than metal lines in the ISM at T 
$\lesssim$ 5000 K.  However, as discussed in $\S$ 
\ref{sect:mnu}, molecular hydrogen may form in the I-fronts of very massive Pop III stars \citep{rs01} and 
significantly raise cooling rates in the neutral gas.  H$_2$ ro-vibrational lines cannot rival metal cooling 
(since they can only lower primordial gas to $\sim$ 200 K instead of 10 - 20 K for metals) but are much more 
effective than H and He alone at temperatures below 5000 K.  H$_2$ formation is greatest in the partially 
ionized gas at the base of the shocked 
shell where energy losses may collapse the shell and trigger dynamical perturbations.  However, the role of H$_2$ 
cooling in primordial I-front instabilities is unclear because massive stars with strong ionizing luminosities 
are also intense sources of Lyman-Werner (LW) photons that dissociate H$_2$.  With energies below the ionization 
limit, these photons can cross the front and degrade the molecular hydrogen catalyzed in its outer layers.  

We formulated three sets of one-dimensional calculations to examine I-front shocks when H, He and H$_2$ cooling 
are present and absent.  Collapse of the shell in one dimension is a rough predictor of instability formation in 
three dimensions.  The initial density profile was: \vspace{0.1in}
\[ n(r) = \left\{ \begin{array}{ll}
			   n_{c}                 & \mbox{if $r \leq r_{c}$} \\
			   n_{c}(r/r_{c})^{-2}   & \mbox{if $r \geq r_{c}$}
                          \end{array}
                  \right.\vspace{0.1in} \] \label{ngas}
\noindent 
with n$_c$ and r$_c$ equal to 1 $\times$ 10$^4$ cm$^{-3}$ and 0.2 pc, respectively.   
  
A UV source with ${\dot{n}}_{ph}$ $=$ 1.0 $\times$ 10$^{48}$ s$^{-1}$ was centered in the mesh, ensuring the 
front would become D-type within the core.  The grid was discretized into 200 uniform radial zones with inner 
and outer boundaries of 0.01 pc and 2.0 pc and reflecting and outflow boundary conditions, respectively.  The 
gas was set to 500 K, appropriate to star-forming minihalos cooled by molecular hydrogen at $z \sim$ 20 
\citep{abn00,abn02}.  We chose a cooling cutoff (defined to be the gas temperature below which cooling curves 
were not applied to gas energy updates) equal to the cosmic microwave background T$_{CMB}$ $=$ 2.73 (1$+$z) K 
with z = 20.  

\begin{figure*}
\epsscale{1.1}
\plotone{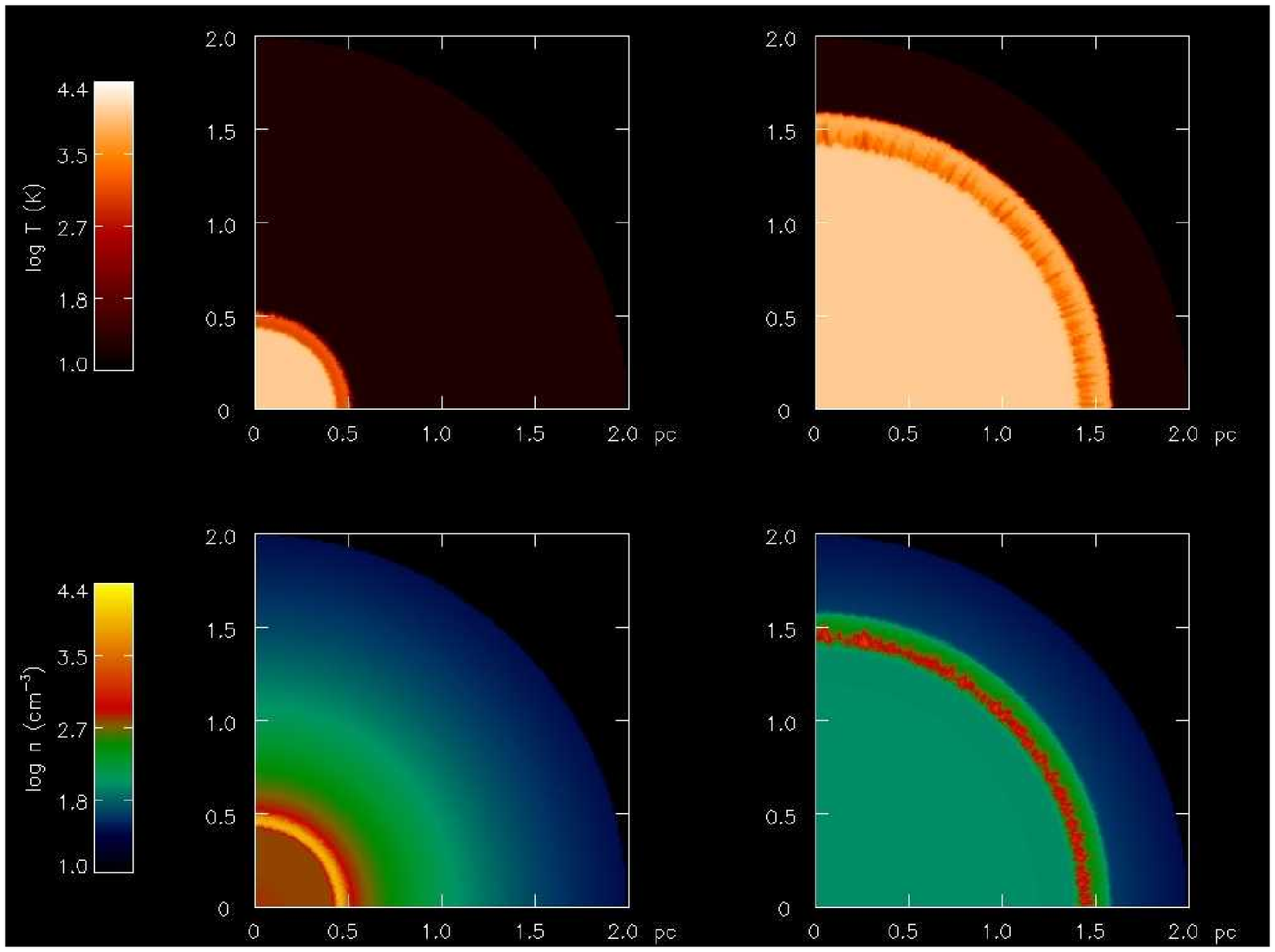}\vspace{0.15in}
\caption{Evolution of a D-type front in an primordial r$^{-2.7}$ envelope.  Panels (a) and (b):  temperatures
at 47.6 kyr and 123.7 kyr, respectively.  Panels (c) and (d):  densities at 47.6 kyr and 123.7 kyr, respectively.
\label{fig:2.7}} 
\vspace{0.075in}
\end{figure*}

All three runs were multifrequency, with blackbody spectra and nine-species primordial gas chemistry 
\citep{anet97}.  The gas profile was initialized with H and He number densities whose sum was equal to n$_c$ 
but 76\% and 24\% H and He by mass, respectively.  We adopted initial electron and H$_2$ number fractions of 
1 $\times$ 10$^{-4}$ and 2 $\times$ 10$^{-6}$ consistent with $z \sim$ 20 \citep{rs01}.  When H$_2$ cooling 
was included, the curves of \citet{gp98} were utilized.  The ionizing photon emission rate corresponds 
approximately to a star with a surface temperature, mass, and luminosity of 57,000 K, 15 M$_\odot$ and 2.1 
$\times$ 10$^4$ L$_\odot$ \citep{s02}, respectively.  We applied these parameters to normalize the photon 
rates binned in each frequency interval.  

\subsection{H/He Cooling in Ionized and Shocked Gas} \label{3.1}

H$_2$ was allowed to form but not cool the gas in the first run.  As an experiment, we toggled the five atomic 
H and He cooling processes to determine their effect on the temperature of the ionized gas.  We found (as 
expected) recombinational cooling to be dominant, reducing the postfront temperature from 60,000 K to 30,000 K.  
The remaining four channels collectively cooled the gas another 4000 K, down to 26,000 K.  All five rates are 
shown in in Fig. \ref{fig:cool} at 100 Kyr in the ionized and shocked neutral gas.  The I-front ends at 0.94 pc 
in the plot and the shocked shell extends to 1.19 pc.  While collisional ionizations briefly surpass 
recombinations near the front, bremsstrahlung, interestingly, is the second most important cooling mode in most 
the of the H II region, primarily due to the relatively large temperatures associated with He ionization.  
Collisional excitation of neutral hydrogen then follows collisional ionization in order of importance in the 
postfront gas, with Compton cooling as the least effective.

Bremsstrahlung and collisional ionization and excitation drop sharply in the shocked gas, leaving only Compton
and recombinational cooling to operate in the low free electron fractions there.  For the sake of comparison, 
we overlay the \citet{dm72} cooling rates for an electron fraction of 0.01 used in \citet{wn07}.  Metal line 
cooling is eight orders of magnitude greater than recombinations or Compton scattering.  Indeed, if no cooling 
cutoff whatsoever was employed with the metal lines they would quickly collapse the shocked gas to very dense
cold shell whose final thickness is bounded from below only by numerical resolution.  We note that the Compton 
component is highly dependent on redshift; had this simulation been performed at z = 35, inverse Compton 
scattering would have dominated recombinations in the postshock shell.  

\begin{figure*}
\epsscale{1.15}
\plottwo{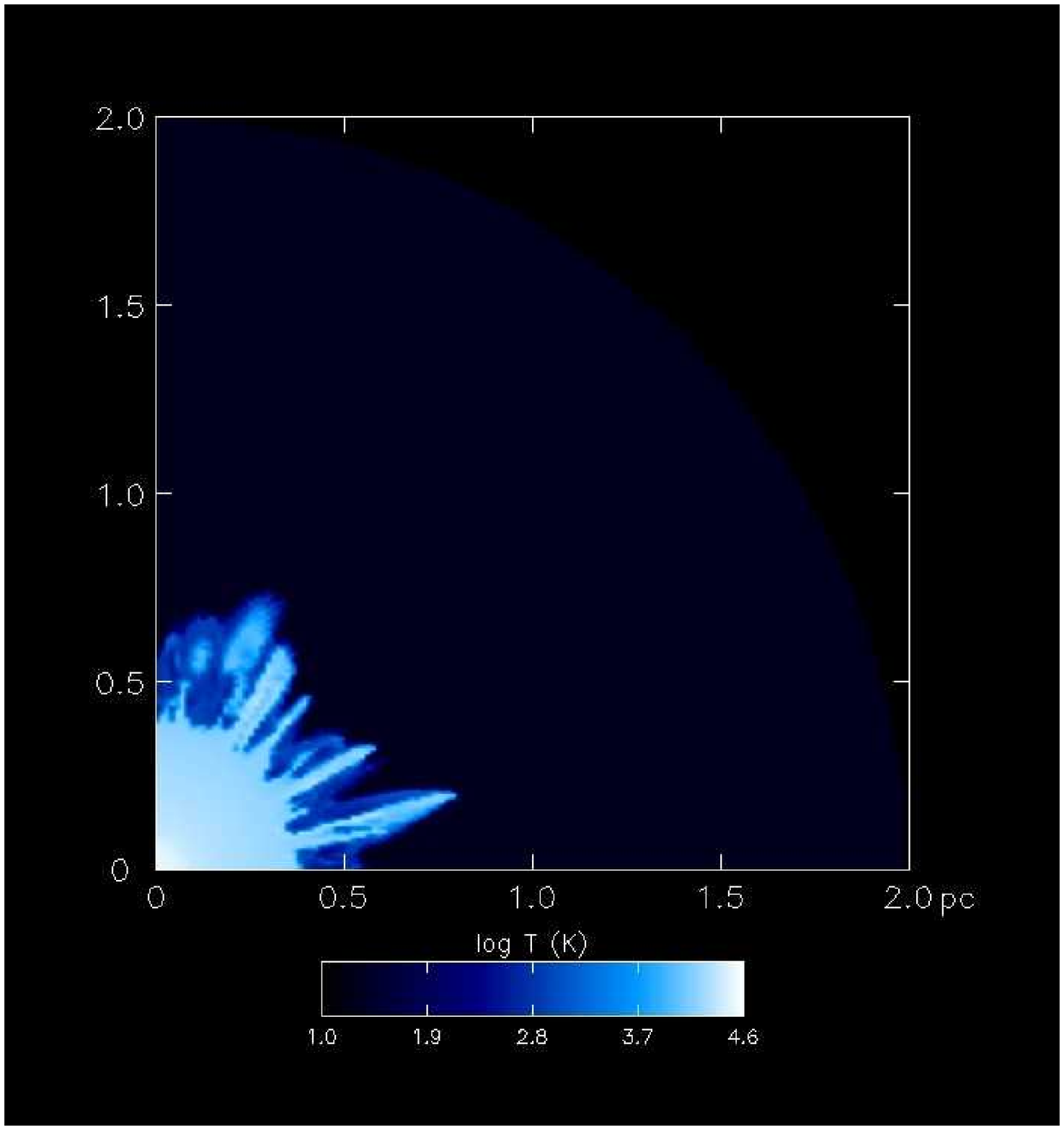}{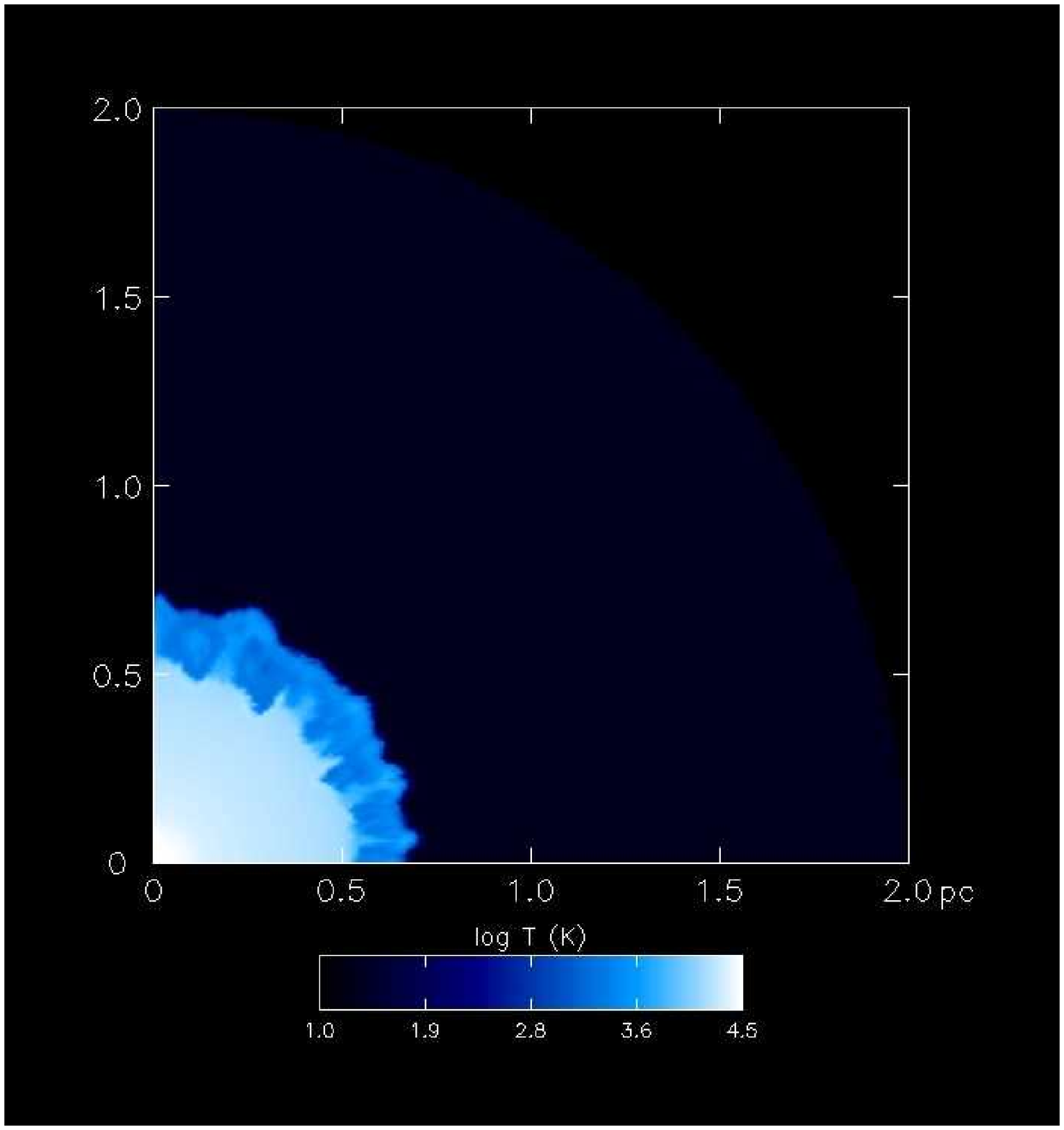}
\caption{Temperature images of H$_2$ mediated thin-shell instabilities in primordial ionization fronts.  
Left: no LW radiation, 40 kyr.  Right: LW radiation included, 51.4 kyr. 
\label{fig:h2}}  
\vspace{0.1in}
\end{figure*}

Recombinational and Compton cooling are so low in the shocked gas that they are inconsequential to its 
structure.  The density and temperature profiles of the shell are indistinguishable from those of a 
nonradiating shock.  Instability formation, except possibly those due to Rayleigh-Taylor perturbations,
is extremely unlikely in these circumstances.

\subsection{H$_2$ Cooling} \label{3.2}

\begin{figure}
\resizebox{3.45in}{!}{\includegraphics{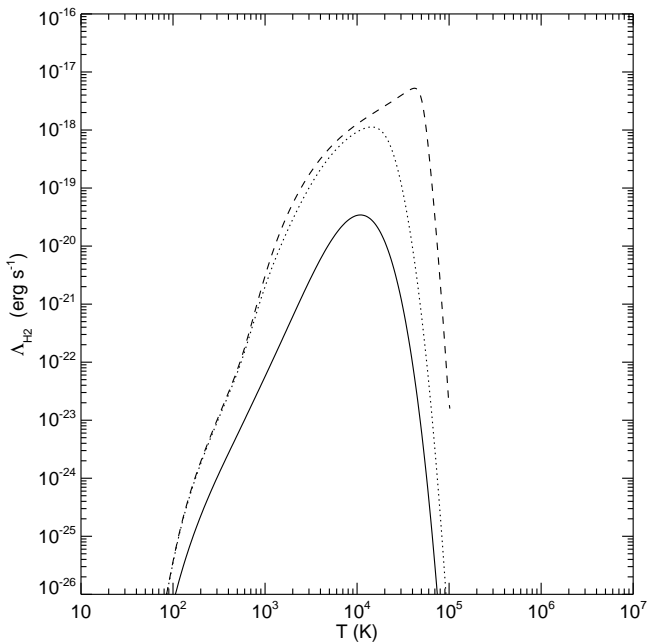}}
\caption{\citet{gp98} molecular hydrogen cooling rates for three fiducial densities. Solid: n$_{H_2}$ 
$=$ 1 $\times$ 10$^{2}$ cm$^{-3}$; dotted: n$_{H_2}$ $=$ 1 $\times$ 10$^{4}$ cm$^{-3}$; dashed: n$_{H_2}$ 
$=$ 1 $\times$ 10$^{8}$ cm$^{-3}$.} \vspace{0.25in}
\label{fig:h2rates}
\end{figure}

In the second model H, He and molecular hydrogen cooling is included but not Lyman-Werner dissociation, so 
H$_2$ forms freely in the outer layers of the front.  In the third model we activate H$_2$ photdissociation
as described in $\S$ \ref{sect:mnu}.  H$_2$ cooling rates, temperatures, densities, and velocities for both 
runs appear in Fig. \ref{fig:cool}.  Molecular hydrogen efficiently drops the base of the shocked shell to 700 
K in the absence of dissociating radiation, condensing the shell to higher peak densities than in the first 
set.  It is likely in these circumstances that H$_2$ cooling would cause unstable modes and shell breakup 
in primordial I-fronts.  

However, at this proximity to the central star molecular hydrogen does not fully self-shield from the 
dissociating UV radiation, as shown in our third model.  The sharp reduction in cooling rates attests to the 
efficiency with which LW photons destroy the H$_2$ molecules (lowering their abundance by more than two orders 
of magnitude), resulting in a shell structure indistinguishable from one with atomic H cooling only.  Comparison 
of the density profiles indicates that H$_2$ cooling reduces the shell thickness by half in the absence of LW 
radiation and that it approaches the efficiency the Dalgarno \& McCray curves within a thin layer at the base.
The velocity profiles and positions of the two fronts indicates that H$_2$ cooling slows the front by 10\% in the 
absence of LW flux.  Nevertheless, even with photodissociation the H$_2$ cooling rates are comparable to the 
Dalgarno \& McCray rates at electron fractions of 0.0001 - 0.001, suggesting that even low molecular hydrogen 
abundances set by LW fluxes may induce unstable modes in primordial I-fronts \citep{wn07}, an issue to which
we return below.

\section{Dynamical Instabilities in D-type Primordial I-Fronts} 

Can atomic cooling in H and He alone incite dynamical instabilities in D-type primordial ionization fronts?
In this section we address both Vishniac and Rayleigh-Taylor instabilities driven by ionization fronts in 
perturbed spherically-symmetric density gradients.  

\subsection{Vishniac Modes} \label{4.1}

Our one-dimensional model in $\S$ \ref{3.1} revealed that I-front shocks cooled only by atomic H and 
He were indistinguishable from nonradiating shocks and therefore are not susceptible to Vishniac 
instabilities.  To verify this prediction we performed a three-dimensional simulation with monoenergetic 
photons to prevent any complications due to multifrequency broadening of the front.  In this run the same 
central photon rate, radial mesh and density profile were used as before but with 180 zones in theta and in 
phi, with reflecting boundaries at $\pi$/4 and 3$\pi$/4 in both angles.  The initial gas and cooling cutoff 
temperatures were set to the CMB background at z = 20 and to 1000 K, respectively.  The energy per ionization 
$\epsilon_{\Gamma}$ was chosen to be 1.6 eV in order to set the postfront temperature to 10000 K.  Thus, while 
He was present in the calculations to preserve its inertia to hydrodynamical motion, it was not ionized.  
Ignoring He ionization simply results in lower ionized gas temperatures, which do not deter instabilities from 
forming (He contributes even less to I-front shock cooling than H in these regimes).

To seed the formation of instabilities, we randomly varied the gas density of each cell by at most 1\%, holding
gas energies constant to prevent pressure fluctuations from prematurely smoothing the perturbations.  We only 
imposed these variations beyond radii of 0.125 pc to prevent the onset of shadowing instabilities in the R-type 
front before dynamical instabilities in the D-type front could form.  In Fig. \ref{fig:S31PM}, the temperatures 
and densities reveal small fluctuations in the ionization and shock fronts due to the original perturbations, 
but they do not grow with time.  The shell simply becomes thicker as more shocked neutral gas accumulates. 
We note the layer of gas photoevaporating into the interior of the H II region from the inner face of the 
shell in panel (a) of Fig. \ref{fig:S31PM}.  It is visible as the thin red arc of lower densities bounding the 
yellow layer of denser shocked gas from below and can only be seen in the absence of instabilities.  Our 
calculation confirms that dynamical instabilities in I-fronts cannot arise through the Vishniac instability in shocked 
gas cooled by only by H and He.  Since our result depends only on cooling efficiencies, they will hold for the 
wide range of densities encountered in primordial halos and protogalaxies.

\subsection{Rayleigh-Taylor Modes}

\begin{figure*}
\epsscale{1.0}
\plotone{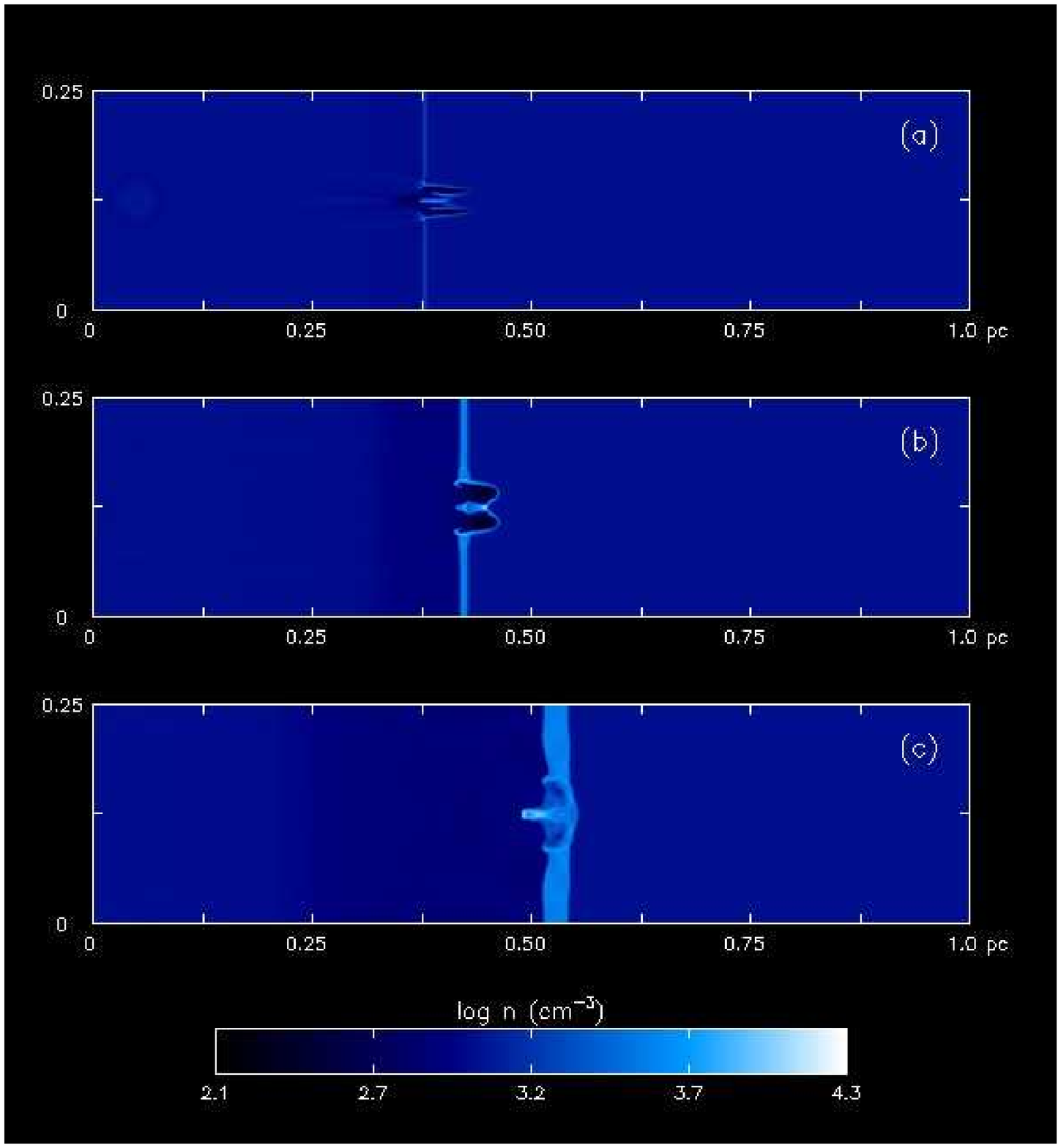}\vspace{0.15in}
\caption{Density images of a shadow instability due to an underdense perturbation in atomic H with h$\nu =$ 
14.2 eV and no H$_2$ cooling: (a) 1.9 kyr, (b) 4.9 kyr, and (c) 12.6 kyr.  \label{fig:rwm}} 
\vspace{0.075in}
\end{figure*}

\begin{figure*}
\epsscale{1.0}
\plotone{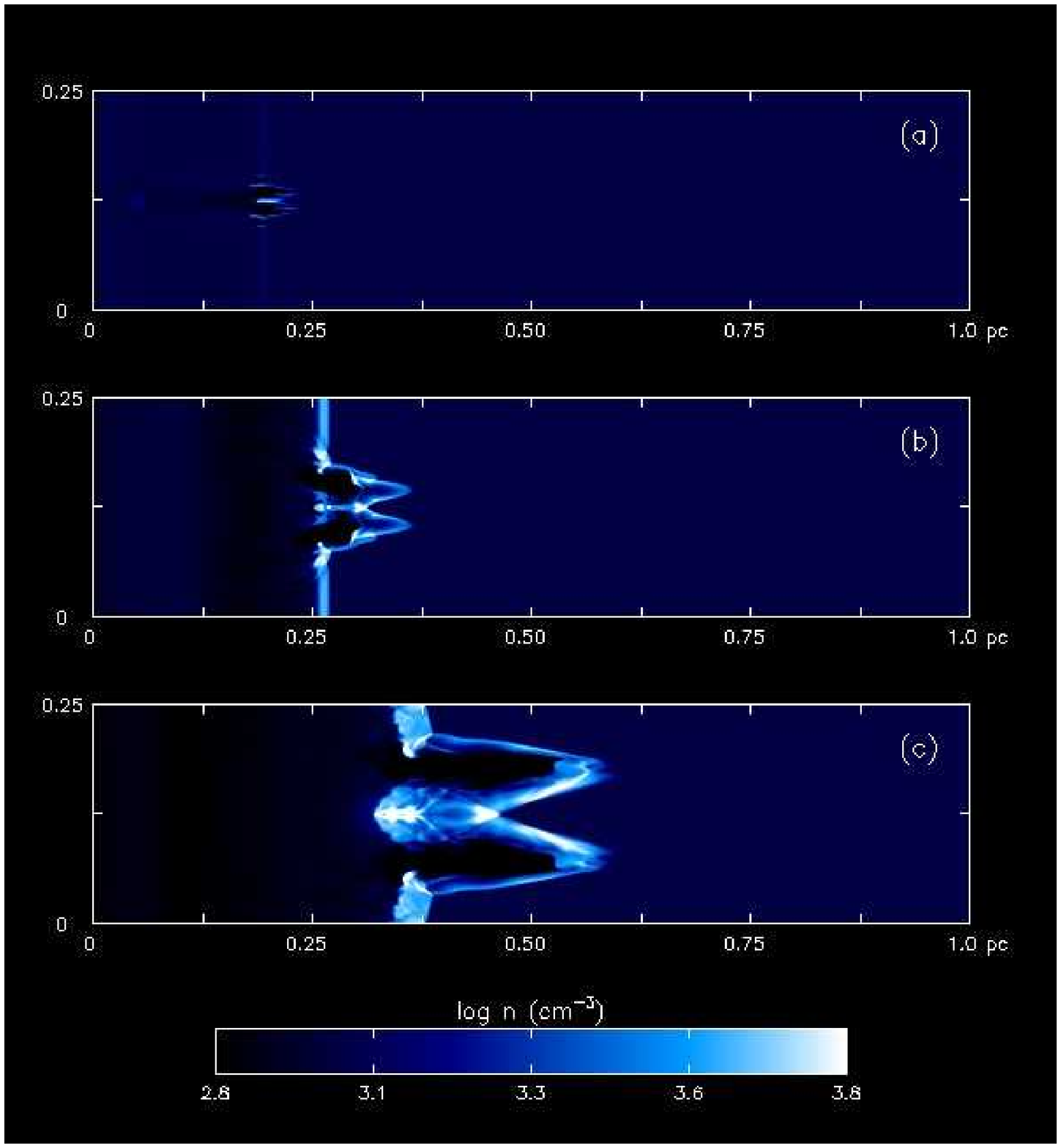}\vspace{0.15in}
\caption{Shadow instability due to an underdense perturbation, full nine species primordial chemistry
with a 100,000 K blackbody spectrum: (a) 1.3 kyr, (b) 6.3 kyr, and (c) 13.4 kyr.\label{fig:rwm_mnu}} 
\vspace{0.075in}
\end{figure*}

Rayleigh-Taylor (RT) instabilities can grow in primordial I-fronts if they accelerate in a 
density gradient, which is the case for profiles steeper than r$^{-2}$.  We performed a run with parameters 
identical to those in $\S$ \ref{4.1} except in a density gradient of r$^{-2.7}$ and show $\phi$ $=$ $\pi$/4 
temperature and density slices of its evolution in Fig. \ref{fig:2.7}.  Perturbations in the dense shell are 
somewhat more pronounced than in the Vishniac run, hinting at the onset of unstable fingers, but they never 
fully materialize and disrupt the shock as the front exits the envelope.  The shock clearly gains strength 
down the gradient, reaching temperatures of nearly 10,000 K.  In such regimes electron-neutral collisions 
soon ionize the shock, helping the front to break through.  While possible in general, we find that RT 
instabilities fail to induce unstable modes in the front at these densities, which are typical of those in 
1 $\times$ 10$^5$ - 1 $\times$ 10$^6$ M$_\odot$ minihalos at early epochs and of molecular cloud cores in 
protogalaxies at later times.

\subsection{Thin-Shell Modes Due to H$_2$ Cooling}

We show in Fig. \ref{fig:h2} the evolution of thin-shell instabilities in primordial D-type ionization fronts
due to molecular hydrogen catalyzed at the base of the shocked shell.  The first run was identical to 
that of $\S$ \ref{4.1} except that H$_2$ cooling was activated, but not LW photodissociation.  The second was 
identical except Lyman-Werner flux was included.  These models span the range of self-shielding anticipated in 
cosmological halos and protogalaxies, from virtually none to complete.  In the absence of LW photons, H$_2$ 
abundances of 1 $\times$ 10$^{-3}$ to 1 $\times$ 10$^{-2}$ form at the I-front/shock interface and, as shown 
in the left panel of Fig. \ref{fig:h2}, are as effective as metals at instigating dynamical instabilities in 
the D-type front (compare with Fig. 5 of \citet{wn07}).  When LW flux is included, equilibrium H$_2$ fractions in the shell fall to 1 $\times$ 
10$^{-6}$ to 1 $\times$ 10$^{-5}$ but, surprisingly, strong instabilities persist despite the diminished cooling.  
The key to this phenomenon is the temperature dependence of the H$_2$ cooling curves, shown in Fig. 
\ref{fig:h2rates}.  From 300 K, the typical temperature of a star-forming cosmological minihalo, to 3000 K, 
the temperature of the shock in the primordial front, H$_2$ cooling rates rise by three orders of magnitude, 
offsetting losses due to the photodissociation of the shell.  We find that the instabilities are not as violent 
but still heavily perturb the shock.

Since our assumed central densities are lower than those in which Pop III stars form, we expect that H$_2$ 
will be more heavily shielded and thin-shell instabilities will be more violent in primordial halos 
and protogalaxies than in this second case.  Preliminary calculations in 1 $\times$ 10$^{6}$ M$_\odot$ 
halo profiles hosting 120 M$_\odot$ stars confirm this suspicion, even with their much larger LW fluxes.  
These two cases likely bracket the magnitudes of the instabilities that occurred in primordial D-type 
ionization fronts.

\section{Shadow Instabilities in R-type Primordial I-Fronts}

To test if the shadow instability can form in primordial, we performed a calculation in which a 
plane-parallel R-type ionization front enters the yz-face of a rectangular box with a uniform density 
except for a spherical underdensity that is slightly offset from the face of entry.  Its density 
decreases linearly in radius from the ambient value at its center to 50\% below this amount at its 
surface.  Our simulation mesh is 1.0 pc along the x-axis and 0.25 pc along the y and z-axes, with 
1000, 250, and 250 zones in the x, y, and z directions respectively.  The gas density in the box is 
1000 cm$^{-3}$ with a temperature of 72 K to establish a sound speed c$_{s}$ of 1 km/s.  The incident 
photon flux along the x-axis at the face of entry is 3.0 $\times$ 10$^{11}$ cm$^{-2}$ s$^{-1}$ to 
approximate that of a 120 M$_\odot$ star at 1.71 pc.  The radius of the perturbation is 0.0125 pc and 
is placed 0.05 pc along the x-axis and centered in the yz-plane.  Inflow and outflow boundary conditions 
were assigned to the 0 pc and 1.0 pc faces, respectively, with reflecting boundaries on the other four 
faces.

\subsection{Monoenergetic Spectrum/Atomic Cooling}

We first consider shadow instabilities in H gas with monoenergetic photons to avoid front broadening, with 
fixed energy per ionization $\epsilon_{\Gamma} =$ 0.8 eV to ensure postfront temperatures of 10000 K.  We 
show xy density slices at z = 0 of the instability in Fig. \ref{fig:rwm} at 1.9 kyr, 4.9 kyr, and 12.6 kyr.  
By 1.9 kyr the front has become D-type and the original corrugation in the R-type front is now nonlinear.  
The expanding photoevaporated perturbation is faintly visible on the left in the figure.  The double peaks 
are cross sections of a ringed jet whose morphology is due to the radial density profile of the sphere.  
The R-type front preferentially advances along lines of sight parallel to the x-axis that cut the 
underdense regions close to the surface of the sphere.  Along the axis piercing the sphere through its center, 
the front advances at nearly the same rate as in the unperturbed medium because the densities along this 
path through the sphere are closer to those in the surrounding gas. 

The evolution of the instability is qualitatively similar at first to that in Fig. 8 of \citet{wn07}, 
but it quickly dampens and becomes subsumed in the inefficiently cooled shocked shell.  
The density knots in the shock located above and below the midplane of the jets at their base in the 
\citet{wn07} run are also visible in this run at first but they rapidly dissipate with little lateral 
migration across the face of the shock.  If the assertion by \citet{rjw99} that these knots are odd-even 
numerical instabilities is true, they cannot survive long in the steadily thickening shocked layer at 
these densities.  Rather than condensing and fragmenting, the shell tends to expand back into the jet, 
crowding its rarefied interior with gas that shields its tip from incident radiation and collapsing it 
back into the face of the shock.  The collapse of the jet back into a dimple imparts considerable 
vorticity to the shocked gas in the vicinity of the disturbance, as is evident in panel (c).  We find 
that atomic hydrogen cooling effectively quenches the shadow instability, but not before it significantly 
churns the face of the shock.

\subsection{Blackbody Spectrum/H$_2$ Cooling}

The jet instability is far more violent when full nine-species primordial chemistry, the blackbody 
spectrum of the 100,000 K primordial star, and molecular hydrogen cooling are included, as we show in 
Fig. \ref{fig:rwm_mnu}.  Here, it is the higher post-front gas temperatures associated with multifrequency 
ionization of H and He that sustain and destabilize the jet.  At 35,000 K, the high pressure of the ionized 
gas prevents the shocked shell ($\sim$ 10,000 K) from expanding and backfilling the rarefied jet, so the dimple
never heals or collapses back into the face of the shock.  The role of H$_2$ cooling is minimal.  The 
thickness of the shell in this model is indistinguishable from that with atomic cooling only, but the
persistence of some molecular hydrogen likely promotes some small-scale fragmentation of the shell.  We observe no 
saturation of the jet and it continues to grow with time.  Maximum overdensities due to clumping of gas
are approximately 10 in both runs.

\section{Metallicity Threshold for D-Type Instability Formation}

\begin{figure*}
\epsscale{1.0}
\plotone{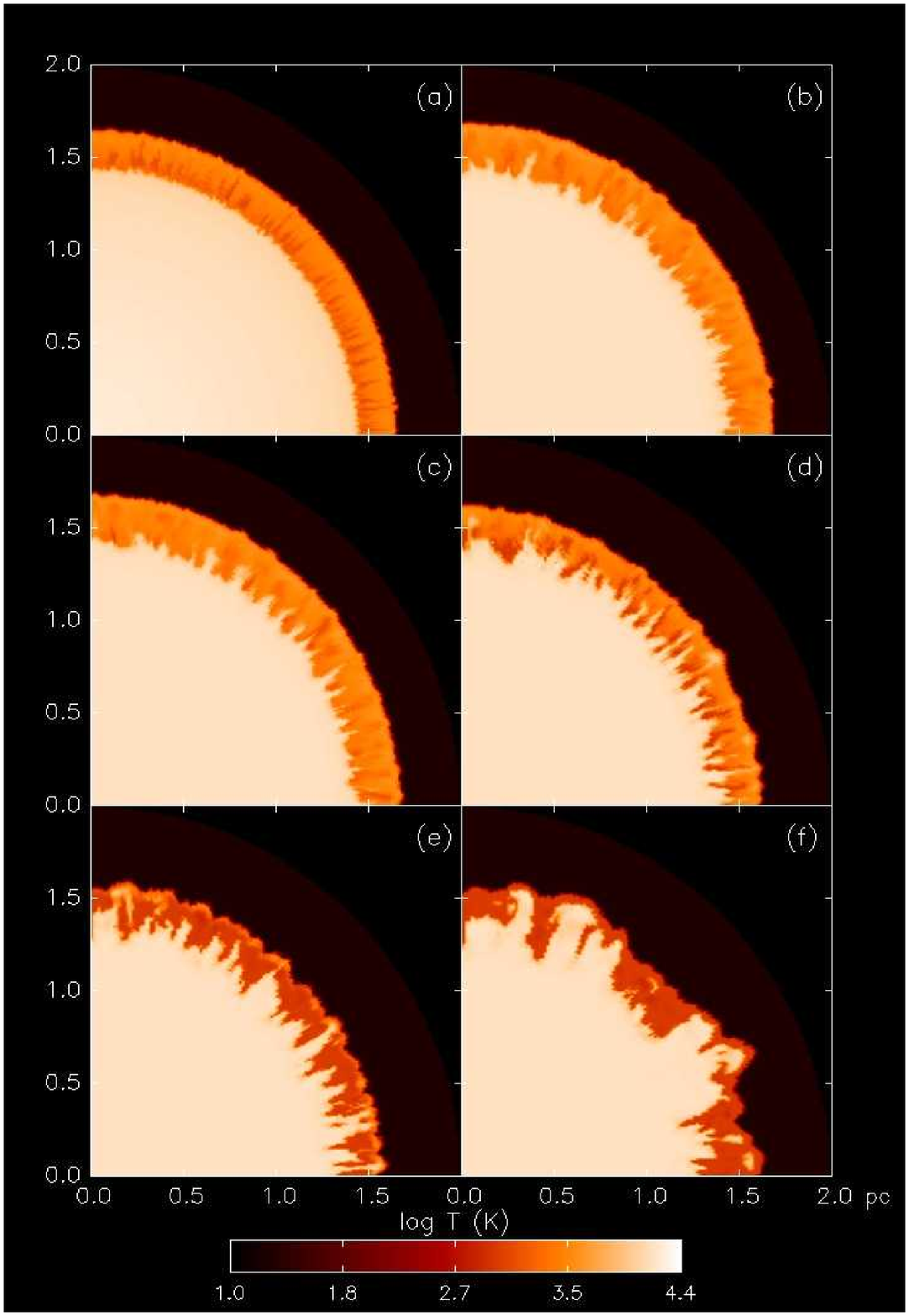}\vspace{0.15in}
\caption{Dynamical instability in an r$^{-2}$ density gradient in a variety of metallicities at 125.5 kyr.  
Panel (a): zero metallicity.  Panels (b), (c), and (d) are for metallicities of 1 $\times$ 10$^{-4}$ Z$_\odot$,
1 $\times$ 10$^{-3}$ Z$_\odot$, and 1 $\times$ 10$^{-2}$ Z$_\odot$, respectively, while panels (e) and (f) are 
for 0.1 Z$_\odot$ and 1.0 Z$_\odot$, respectively.\label{fig:metallicity}} 
\vspace{0.075in}
\end{figure*}
In $\S$ 4 it was shown that violent dynamical instabilities do not manifest in shocked shells cooled by 
atomic H only.  At what metallicity do dynamical instabilities appear in D-type I-fronts in the absence 
of H$_2$ cooling?  We performed a grid of calculations with Dalgarno \& McCray cooling in metallicities 
of 1 $\times$ 10$^{-4}$ Z$_\odot$, 1 $\times$ 10$^{-3}$ Z$_\odot$, 1 $\times$ 10$^{-2}$ Z$_\odot$, 0.1 
Z$_\odot$, and 1.0  Z$_\odot$ to ascertain if there is threshold to instability growth.  The problem 
setup is the same as in the three-dimensional r$^{-2}$ run described in $\S$ \ref{4.1}, again assuming 
an initial electron fraction of 1 $\times$ 10$^{-4}$ appropriate for z $\sim$ 20.  At 
such low fractions the cooling in the gas is chiefly due to excitation of metals by collisions with 
neutral hydrogen (see Fig. 2 in \citet{dm72}).  At solar metallicity these cooling rates lie 
approximately an order of magnitude below the Dalgarno \& McCray curve shown in Fig. \ref{fig:cool}.  
Our metallicities were chosen to survey cooling efficiencies between the very low purely atomic rates 
and quite large 1 Z$_\odot$ metal line curve.  We apply a cooling cutoff temperature of 1000 K rather 
than the CMB temperature. This limits the thickness to which the shell can collapse so the metallicity 
at which instabilities appear is truly the threshold for their formation.

In Fig. \ref{fig:metallicity} the I-front and shock are shown at 125.5 kyr for our six choices of metallicity.
We include in panel (a) the front in a zero-metallicity gas as a reference case in which no unstable modes
arise.  Noticeable perturbations appear in the front in roughly equal numbers in gas enriched to 1 $\times$ 10$^{-4}$ 
Z$_\odot$ and 1 $\times$ 10$^{-3}$ Z$_\odot$ (panels (b) and (c), respectively); those in panel (c) are slightly more
pronounced but in neither case do they grow significantly with time or penetrate the shell to disrupt the
shock.  The agitation in the front does perturb the shock in the 1 $\times$ 10$^{-2}$ Z$_\odot$ gas, with visible
hydrodynamical motions above and below the image plane.  Also, consolidation of initially short-wavelength modes
into fewer larger-wavelength structures is evident, clearly demonstrating the evolution in D-type fronts
predicted by \citet{gu79}.  These features are even more pronounced in the 0.1 Z$_\odot$ case, and the 
instability completely deforms the shock in the 1.0 Z$_\odot$ gas.  

If one adopts deformation of the shock and wavelength evolution of the unstable modes as the criteria for
instability formation, it appears that metallicities of $\sim$ 1 $\times$ 10$^{-2}$ Z$_\odot$ are the threshold for
perturbation growth in these densities.
However, we note that primordial gas in relic H II regions would be susceptible to ionization front 
instabilities at considerably lower metallicities because they are compensated by larger free electron 
abundances.  Our results indicate that this type of flow instability would operate even in marginally 
enriched protgalaxies and star-forming minihalos.  The morphology of the instabilities at later times
suggest that they might be efficient at driving turbulence in the shock, which could have important 
implications for the time scales of subsequent star formation in the region \citep{wn07}.  Finally, we 
note that instabilities readily erupt at solar metallicities even with the much lower initial electron
fraction $\chi_e$ of 0.0001 than the $\chi_e$ = 0.01 value adopted in \citet{wn07}.  While in general
it is difficult to know what free electron abundances are present in a galactic molecular cloud, a
reasonable assumption is that all the carbon in the gas is singly ionized by the UV background.  This
yields a lower bound for electron fractions in star forming regions in the galaxy today that is on the 
order of those applied in this grid of metallicities, confirming that I-front instabilities will be a
general feature of those regions. 
 
\section{Discussion and Conclusions}

Instabilities readily form in primordial ionization fronts, even in the absence of metals.  
However, they emerge only when calculations couple multifrequency radiative transfer to gas phase 
primordial chemistry, which is crucial to capturing the formation of H$_2$ that triggers dynamical 
instabilities in self-shielded shells or the high post-front gas temperatures that perpetuate 
shadow instabilities.  We find that thin-shell modes cannot develop in atomic H and He but that
Rayleigh-Taylor instabilities are not ruled out.  Nevertheless, they do not appear on the time 
scales of UV breakout from the density profiles in our models, which roughly approximate those in 
cosmological halos.

Dynamical instabilities in D-type fronts were probably common in centrally-concentrated structures 
capable of molecular hydrogen self-shielding, like cosmological minihalos at subparsec radii and 
molecular cores in primitive galaxies.  However, shadow instabilities appearing earlier in R-type 
fronts and needing no H$_2$ cooling may have pre-empted thin-shell instabilities in more evolved 
I-fronts.  Inhomogeneities endemic to all primordial structures made this instability ubiquitous, 
forming shocks and clumps that regulated the escape of ionizing UV photons and metals from the first 
luminous objects.  

One question that remains to be answered is how instabilities forming at subparsec radii deep in 
cosmological halos or early galaxies governed the final morphologies of H II regions on kiloparsec
scales.  If clumping due to unstable modes was long-lived, UV breakout along adjacent lines of
sight may have been significantly delayed, causing the formation of extended fingers of ionized
gas into the early IGM that resulted in serious departures from the classical ``butterfly'' appearance 
of many cosmological H II regions \citep{awb06}.  Numerical models bridging the spatial scales 
between the young D-type front and final H II region are now under development.

Ionization front instabilities in the accretion envelopes of the first stars may have allowed them 
to grow to much larger masses than predicted by analytical models of infall reversal \citep{tm04}.
Fingers of ionized gas elongating outward through the accretion shock might have channelled ionizing 
radiation out of the cloud that would otherwise have staunched infall.  Accretion could then continue 
downward in columns of neutral flow, allowing the star to bypass the final mass limits set by the
cutoff models.    

The strength and longevity of I-front instabilities determined their impact on subsequent star 
formation.  Prominent extrusions from the front out into the halo or host galaxy might have created
clumps of primordial gas prone to gravitational collapse by HD cooling in the relic H II region,
forming a new generation of stars without recourse to chemical enrichment.  This scenario merits 
further study with high resolution methods, 
as it is uncertain if clumps could have collapsed outside the potential well of the dark 
matter halo.  On the other hand, more saturated modes due to less efficient cooling and fragmentation 
of the shell may instead have provided turbulent support against condensation in larger cloud complexes within 
protogalaxies, suppressing rather than promoting star formation. 

Clumping by instabilities may have determined the extent of the first metal bubbles \citep{jet07} in 
addition to modulating UV escape fractions and the final radii of H II regions.  Mixing of metals with 
clumps may also have accelerated second generation star formation.  Efficiently cooled clumps would
have collapsed to form stars on mass scales very different from their predecessors.  Accurate models of primordial 
H II regions that include ionization front instabilities will establish the proper initial conditions 
for metal mixing in the next generation of early chemical enrichment studies.  

Although dynamical instabilities due to heavier elements appeared at metallicities of 0.001 - 0.01 at 
ionized fractions of 0.01, they would have appeared in abundances an order of magnitude lower in the
higher free electron fractions of fossil H II regions.  This is comparable to the metallicity associated 
with the transition from Pop III stars to less massive populations \citep{mbh03}; hence, instability
formation, and its consequences, smoothly followed rollovers in initial mass functions (IMFs) in the 
early universe.

We neglect direct transport of recombination photons, which may limit the growth of unstable
modes.  Reprocessed radiation emanating from within ionized extensions of gas may erode the sides of
the fingers, photoevaporating overdensities in the troughs.  This may not be an important effect in
shadow instabilities since the interior of the jet is relatively rarefied, with low recombination
emissivities.

\acknowledgments

DW thanks Tom Abel, Kyungjin Ahn and Alex Heger for helpful discussions concerning these simulations and
we thank the anonymous referee whose comments improved the quality of this paper.  This work was carried 
out under the auspices of the National Nuclear Security Administration of the U.S. Department of Energy 
at Los Alamos National Laboratory under Contract No. DE-AC52-06NA25396.  The simulations were performed 
at SDSC and NCSA under NRAC allocation MCA98N020.

\end{document}